\magnification=\magstep1
\hfuzz=16pt
\baselineskip = 17pt
$ $

\vskip 1in

\centerline{\bf Analysis of a work of quantum art}

\bigskip

\centerline{Seth Lloyd}

\centerline{Department of Mechanical Engineering}

\centerline{Massachusetts Institute of Technology}

\centerline{MIT 3-160, Cambridge MA 02139 USA.  slloyd@mit.edu}

\bigskip

\noindent{\it Abstract:}  This paper provides a quantum-mechanical analysis
of an artwork, `Wigner's friends,' by Diemut Strebe.   The work consists of
two telescopes, one on earth, one launched into space, and explores ideas
of quantum correlations and quantum measurement.   This paper examines the
scientific basis of the work and analyzes
the form of quantum correlation between the two telescope systems.

\vskip 1cm 

On November 23rd 2014, the Italian astronaut Samantha Cristoforetti launched
a small telescope from the International Space Station.    The twin
of this telescope remained on earth, and will be attached to the
James Webb telescope when it is launched in 2018.  The two telescopes
are part of an artwork by Diemut Strebe, entitled `Wigner's friends.'
The title of the work refers to Eugene Wigner's exploration of the
quantum measurement problem in [1].   
In the ordinary quantum time evolution governed by the
Schr\"odinger equation, a measurement made on a quantum 
system that is initially in a superposition of two different states 
causes first the measurement apparatus to become correlated with
the quantum system, and then the brain of the scientist who observes
the measurement apparatus, and so on.  The wave function of the system,
measurement apparatus, and observer contains all possible outcomes
of the measurement.   Wigner noted that 
to obtain a quantum description in
which only one outcome of the measurement is represented in the
quantum wave function requires some additional effect, sometimes
called `wave function collapse,' in which the part of the 
wave function that is not actually measured or perceived to be
the case goes away, leaving only the observed part.  Wigner
asked whether his friend -- before she has told Wigner the result
of her measurement -- is actually in a quantum superposition.
By contrast, if no additional dynamics beyond the Schr\"odinger equation is
invoked, then the wave function of the system, measurement apparatus, 
and observer contains all possible outcomes of the measurement, a
counter-intuitive situation that is sometimes called the `many
worlds' account of quantum measurement [2].     
Strebe's work is a play on the idea of Wigner's friend and the many
worlds theory.     Although the work is a piece of art, not of science,
she bases the work on several claims which do have scientific content.
The purpose of this note is to discuss the scientific content of her
claims.   

Each telescope consists of an aluminum tube 40mm long, with a 9mm 
inner diameter and a 10mm outer diameter.  The outer surface is
anodized violet.  Each telescope is connected
to a commercial cellphone camera with plastic lenses and a 5 megapixel
CCD array.  Until it is launched on the James Webb telescope, the earthbound
telescope will be powered and aimed at the sky.    The telescope 
that was launched from the international space station is unpowered.

The first part of Strebe's work is called `the universal show.'   Following
on the Jorge Luis Borges story, `The Library of Babel,' Strebe notes
that according to Wigner, in the absence of an observer to `collapse' its wave 
function, the telescope in space contains simultaneously all possible artworks 
in quantum superposition in the pixels of its CCD array.  The great 
majority of the images in the superposition will be random, but a small
fraction of them will have patterns.   (Borges's story imagines a library
in which all possible books have been collected, the vast majority of them
completely random.) 
Let's examine this claim in 
the light of various theories of quantum mechanics.  The quantum
state of the telescope's CCD array depends on the object in its field
of view.   For the moment, assume that the telescope is in the shadow
of the earth, and pointing at interstellar space, so that the temperature
of the telescope is the ambient $3K$ temperature of the cosmic
microwave background radiation and the primary electromagnetic radiation
entering the telescope is the same cosmic microwave background radiation.
Note that the because of the dimensions of the telescope, this
radiation enters the telescope through an attenuated, evanescent mode.
(When the telescope is 
pointing at the sun, the conclusions will be similar, except that
the light entering the telescope will now be in the visible range
and its modes
can be spatially resolved by the lens of the telescope and the CCD array.)
This radiation causes the elements of the CCD array to
undergo energy fluctuations corresponding to thermal radiation at $3$ degrees
Kelvin.   Assigning numerical values to these fluctuations, and interpreting
the numerical values as an intensity scale implies that the elements of the
array are indeed registering all possible 5 megapixel images in quantum
superposition.   

It may be objected that the light entering the telescope is in a thermal
state rather than a pure state, and so is technically
a mixture rather than a quantum superposition.  Note, however, that
in the currently accepted quantum models of cosmology, the universe began
in a pure state, underwent a period of rapid inflation, and then thermalized
in the lead up to the big bang.   If the time evolution of the quantum
fields in the universe remains unitary throughout this evolution, then
the universe remains today in a pure state: pieces of the universe such
as the CCD array of the telescope can be in a mixed state, but only as
the result of entanglement with the rest of the universe.   That is,
the thermal fluctuations in the CCD array are part of an entangled
superposition with the remaining matter and energy in the universe.  
Accordingly, it seems not unreasonable to refer to the this fluctuating    
array as containing all possible images in quantum superposition,
a quantum version of Borges' library of Babel.

Of course, if one adopts an environmentally induced decoherence picture
in which interactions between the elements of the CCD array and
their environment effectively
`collapse' the wave function, then one could also reasonably claim
the CCD array to have collapsed to one out all of these possible images.
(Note that the CCD array of the telescope in space is unpowered, and so
is not making an active measurement of the thermal fluctuations in its
elements.   Another difference between the situation where the telescope
is pointing at the sun and at interstellar space is that the $3K$ radiation
would be insufficiently strong to trigger a CCD signal even if the array
were powered up, whereas the light from the sun is would trigger such 
a signal.)    If one adopts either the many worlds interpretation
or the perspective of Wigner, however, then
the CCD array remains in a superposition of all possible images -- Wigner
has no friends in space.

Now consider the second claim of Strebe, that the telescope on the ground
and the telescope in space exhibit quantum correlations that effectively
connect the two telescopes, and that when the telescope on the ground
detects a photon and creates a signal that can be observed by the
viewers of the exhibit, the state of the telescope in space -- conditioned
on the state of the telescope on earth -- undergoes a change
as in Einstein's notion of `spooky action at distance.'   Clearly,
this change, if it exists, must be very slight.  Nonetheless,
the Hanbury Brown Twiss effect [3-6]
shows that a slight correlation between the photons entering
the telescopes does indeed exist, and so detection of a photon by 
the ground based telescope does indeed change by a small amount
the probability that a photon enters the space based telescope.
This correlation is intrinsically quantum mechanical [4-6] and
relies on the same coherent quantum effects that give rise to
photon bunching and anti-bunching.  Applying concepts of quantum
information theory to Fano's elegant treatment of the Hanbury
Brown Twiss effect in the low-photon number limit [4], we
find that the quantum
state of the two telescopes exhibits a non-zero amount of the
form of quantum correlation [7] known as discord [8-9] (more
technically, the density matrix for
the two telescopes possesses off-diagonal terms when expressed
in the Schmidt bases for the individual telescopes).  
Although discord is a weaker form of quantum correlation than 
entanglement, the existence of discord entering the two 
telescopes implies that any measurement made
on one telescope must have an effect on the overall probabilities
of joint measurements made on both.  The telescopes do indeed
possess a small measure of spooky action at a distance.
 
For the purpose
of analysis, suppose that the two telescopes are momentarily focused
on the same distant star.
Light with wavelength $\lambda$ entering the telescopes from the star is 
spatially coherent as long as the telescopes lie within the transverse
coherence length of the source, $\ell_t \approx \lambda R/r$, where
$R$ is the distance to the source and $r$ is the radius of the source [6]. 
For long wavelength light, the transverse coherence length of a distant star can
easily extend for thousands of kilometers, so that the telescope
in space and the telescope on the ground lie within the coherence length.
The Hanbury Brown Twiss coherence length represents classical
coherence of the waves entering the telescopes from the distant
source.  Quantum effects arise when one considers the particle-like
nature of light, and the fact that that photons are indistinguishable
particles with bosonic statistics [4-6]: 
for the two telescopes, the photons arriving from
the star will be bunched -- if a photon arrives at one telescope,
a photon is more likely to arrive at the other telescope.   Because of 
the small ratio of the aperatures of the telescopes to the area of
the region of transverse coherence, the quantum mechanical bunching 
effect is also small.  But small is not zero.
Accordingly, Strebe's claim
that the two telescopes exhibit quantum correlation seems not unreasonable.

If the two telescopes are momentarily focused on a single photon
source such as a fluorescing atom, then the photons entering
into the telescope exhibit anti-bunching [5]. 
Consider a single photon emitted by the source
into a quantum superposition of two 
spatio-temperal modes entering the two telescopes at the same time (to
within an accuracy $\Delta t \approx \lambda/c$). 
Conditioned on the presence of such a photon, the states of the CCD arrays
within the two telescopes are entangled, and detection of a photon
by the CCD array on the ground reduces the probability of a photon
being absorbed by the CCD array in space to zero.  
This entanglement arises under the condition that a single such
photon enters both telescopes simultaneously [5].   Without such a
condition, the telescopes are in highly mixed, almost uncorrelated 
states and are almost certainly not entangled.   Again, however,
the very slight quantum correlation between the telescopes implies
that their CCD arrays typically exhibit a small amount 
of quantum discord. 

To see the existence of discord in a mathematically explicit way,
consider Fano's model of two atoms, both initially in the excited state, that
transfer to their excitations to two other atoms, both initially
unexcited.  Following Fano, we call the emitting atoms $a,b$ and
the absorbing atoms $c,d$.   Let $D_{ac}, D_{ad}$ be the amplitudes that 
$a$ emits a photon that is absorbed by $c,d$ respectively; similarly,
$D_{bc}, D_{bd}$ are the amplitudes for photon transfer from $b$
to $c,d$.    Suppose that emitters $a,b$ are initially in the excited state
and the detectors $c,d$ are initially in their ground state.
The probability that $c,d$ are both excited is
$p_{cd} = |D_{ac} D_{bd} + D_{ad} D_{bc}|^2$.  Note that this amplitude squared
contains quartic cross terms such as $ D_{ac} D_{bd} \bar D_{ad} \bar D_{bc}$
that contribute to the probability.  As long as the two absorbers are
within the HBT coherence length, these cross terms persist even
when the emission is incoherent [4-6].  Comparing this coherent probability
with the incoherent
`probability' $ \tilde p_{cd} = |D_{ac} D_{bd}|^2 + |D_{ad} D_{bc}|^2$ 
that either
$a$ excites $c$ and $b$ excites $d$, or that $a$ excites $d$ and
$b$ excites $c$, we see that coherence -- classical in the case of 
interference between the electromagnetic waves emitted by $a,b$, 
and quantum in the case of single photons --
makes a difference in the probability of joint excitation of $c$ and $d$.
This coherence is the origin of the Hanbury Brown Twiss effect: averaged
over all emitters in a distant star it leads to photon bunching -- an
enhancement of the joint absorption probability.  

Applied to a single
emitter, the same coherence leads to anti-bunching: a single photon emitted
by the emitter can be absorbed by only one of the detectors.  
In the case of anti-bunching, a photon emitted by a single atom leads
to a quantum superposition state 
$|\psi_{cd}\rangle = (1/\sqrt 2)(|1\rangle_c|0\rangle_d 
+ e^{i\phi} |0\rangle_c|1\rangle_d)$, where the phase $\phi$ depends
on the distances from the emitter to the absorbers.  As noted above, such
a state is entangled.  In the actual telescopes, of course, the atoms
will be in a highly mixed state that at any moment is probably not
entangled.   Nonetheless, the very slight admixture of a single
entangled state typically leads to the creation of quantum discord [8-9]:
for example, if the two atoms $cd$ in the two different telescopes are 
in uncorrelated thermal state, with zero discord, and a single photon
in state $|\psi_{cd}\rangle$ is mixed in with non-zero probability $\epsilon$,
then no matter how miniscule $\epsilon$ is, the resulting state possesses
non-zero discord. 
When the joint state of the telescopes exhibits even a small amount of discord,
then a measurement made on one telescope changes the quantum
state of the two telescopes taken together.     
The exact characterization of the set of states such that
admixture of an entangled state leads to quantum discord is
an open question.

\bigskip\noindent{\it Summary:} Quantum mechanics is well-known to be
strange and counterintuitive.  It should come as no surprise that
weird quantum effects have inspired works of art.  This note analyzed
the quantum mechanics behind a recent quantum artwork consisting of
two telescopes, one of which has been launched into space.      
The art evokes Wigner's account of the quantum measurement problem,
the many worlds theory of quantum mechanics, and Einstein's notion
of `spooky action at a distance' ({\it spukhafte Fernwirkung}).
No claim here is made for the artistic merit of `Wigner's friends': 
judgement must remain
with the viewers of the artwork.   However, the science on
which the artwork is based seems to fall within the
bounds of artistic license.

\vfill
\noindent{\it Acknowledments:} The author thanks H.J. Kimble for useful
discussions and references on the Hanbury Brown Twiss effect (while in
no way implying HJK's endorsement of any statements made here).
\vfil\eject

\noindent{\it References:}

\bigskip\noindent [1]
E.P. Wigner, {\it Symmetries and Reflections}, Indiana University Press,
Bloomington, 1967.

\bigskip\noindent [2]
H. Everett, {\it Reviews of Modern Physics} {\bf 29} (3), 454-462 (1957).

\bigskip\noindent [3]
R. Hanbury Brown, R. Q. Twiss, {\it Nature} {\bf 178} (4541), 1046–1048 (1956).

\bigskip\noindent [4]
U. Fano, {\it American Journal of Phhysics} {\bf 29}, 539 (1961).

\bigskip\noindent [5]
H.J. Kimble, M. Dagenais, L. Mandel,  
{\it Physical Review Letters} {\bf 39} (11), 691 (1977).

\bigskip\noindent [6] L. Mandel and E. Wolf, {\it Optical coherence
and quantum optics,} Cambridge University Press, Cambridge 1995.    

\bigskip\noindent [7] M. A. Nielsen and I. L. Chuang,
Quantum Computation and Quantum Information
(Cambridge University Press, 2000).

\bigskip\noindent [8] H. Ollivier and W.H. Zurek,
{\it Phys. Rev. Lett.} {\bf 88}, 017901 (2001).

\bigskip\noindent [9] L. Henderson and V. Vedral, {\it J. Phys. A:
Math. Gen.} {\bf 34}, 6899-6905 (2001).

\vfill\eject\end